\documentclass[twocolumn,prc,superscriptaddress,unsortedaddress,showpacs,preprintnumbers,amsmath,amssymb]{revtex4}

\def\fun#1#2{\lower3.6pt\vbox{\baselineskip0pt\lineskip.9pt
  \ialign{$\mathsurround=0pt#1\hfil##\hfil$\crcr#2\crcr\sim\crcr}}}

\usepackage{graphicx}
\usepackage{dcolumn}
\usepackage{bm}

\begin{document}

\title{Effect of tensor correlations on the depletion of nuclear Fermi sea within the
extended BHF approach}

\author{Peng Yin}
\affiliation{Institute of Modern Physics, Chinese Academy of
Sciences, Lanzhou 730000, China} \affiliation{University
of Chinese Academy of Sciences, Beijing, 100049, China}
\author{Jianmin Dong}\affiliation{Institute of Modern Physics, Chinese Academy of
Sciences, Lanzhou 730000, China}\affiliation{State Key Laboratory of Theoretical Physics,
Institute of Theoretical Physics, Chinese Academy of Sciences, Beijing 100190, China}
\author{Wei Zuo\footnote{Corresponding author: zuowei@impcas.ac.cn}}
\affiliation{Institute of Modern Physics, Chinese Academy of
 Sciences, Lanzhou 730000, China}\affiliation{State Key Laboratory of Theoretical Physics,
 Institute of Theoretical Physics, Chinese Academy of Sciences, Beijing 100190, China}

\begin{abstract}
We have investigated the effect of tensor correlations on the depletion of nuclear Fermi sea in
in symmetric nuclear matter within the framework of the extended Brueckner-Hartree-Fock approach by adopting
the $AV18$ two-body interaction and a microscopic three-body force. The contributions
from various partial wave channels including
the isospin-singlet $T=0$ channel, the isospin-triplet $T=1$ channel and the $T=0$ tensor $^3 SD _1$ channel have been
calculated. The $T=0$ neutron-proton correlations play a dominated role in causing the depletion
of nuclear Fermi sea. The $T=0$ correlation-induced depletion turns out to stem almost completely from the $^3SD_1$ tensor channel.
The isospin-singlet $T=0$ $^3SD_1$ tensor correlations is shown to response for most of the depletion,
which amounts to more than 70 percent out of the total depletion in the density region considered.
The TBF turns out to leads to an enhancement of the depletion at high densities well above the empirical saturation density
and its effect increases as a function of density.

\end{abstract}

 \maketitle

\section{Introduction}
Tensor correlations and short-range correlations, which may lead to the depletion of the hole states and the partial population
of the particle states, play a crucial role in understanding many properties of
nuclear many-body system,
In Ref.\cite{otsuka:2005,otsuka:2010} the tensor component of nucleon-nucleon ($NN$) interactions
has been shown to responsible for the shell evolution of exotic nuclei.
The investigation of Ref.\cite{vidana:2011} indicate that the contribution of the neutron-proton ($np$)
tensor cannel is decisive for determining
the density dependence of nuclear symmetry energy. In dense neutron star matter, the short-range
tensor correlations are expected to be especially important for understanding the cooling mechanism,
the nucleon pairing and the transport phenomena of neutron stars~\cite{frankfurt:2008}.
Due to the $NN$ correlations induced by $NN$ interactions, a nuclear many-body system
behaves much more complicatedly and manifests more plentiful properties than a non-interacting Fermi system.
For example, the $NN$ correlations, especially the tensor correlations, may lead to a depletion of the nucleon momentum
distribution below the Fermi momentum and a partial population above the Fermi momentum in nuclear matter\cite{jeukenne:1976}.
 The depletion of the Fermi sea measures of the strength of the dynamical $NN$ correlations induced by
the $NN$ interaction in a nuclear many-body system~\cite{vonderfecht:1991,cavedon:1982,pandharipande:1997},
and It plays an important role in testing the validity of the physical picture of independent particle motion in
the standard shell model and/or a nuclear mean field model.
Since the depletion of the lowest hole states in nuclear matter can be identified approximately as
the depletion of the deeply bound states inside heavy finite nuclei, investigation of the nucleon momentum distribution
in nuclear matter may provide desirable information for understanding the structure of finite nuclei.

Experimentally, the effects of $NN$ correlations and the nuclear depletion of the Fermi sea can be investigated by
the ($e,e'p$), ($e,e'NN$) and proton induced knockout reactions~\cite{ramos:1989,dickhoff:1992,dickhoff:2004}.
Up to now, the related measurements have been performed at various laboratories
continually~\cite{mitt:1990,lapikas:1999,starink:2000,batenburg:2001,rohe:2004,niyazov:2004,
benmokhtar:2005,aclander:1999,onderwater:1998,reley:2008,subedi:2008} and definite evidence of the
short-range $NN$ correlations has been observed in these experiments.
The ($e,e'p$) experiments on $^{208}$Pb at NIKHEF has indicated that the deeply bound proton
states are depleted by $15\% - 20\%$ in order to explain the measured
coincidence cross sections~\cite{batenburg:2001}. Recent experiments on the
two-nucleon knockout reactions have shown that nucleons can form short-range correlated pairs with large
relative momentum and small center-of-mass momentum. The observed strong enhancement of the $np$ short-range pairs
over the proton-proton($pp$) pairs at JLab~\cite{subedi:2008} has indicated the dominate role played by
the short-range tensor correlations in dense nuclear medium~\cite{schiavilla:2007}.
Theoretically, the nucleon momentum distribution and the short-range correlations in
 nuclear matter have been investigated extensively by using various microscopic approaches, such as
the Green function theory~\cite{muther:1995,alm:1996,dewulf:2003,frick:2005,rios:2009},
the extended Brueckner-Hartree-Fock (BHF) framework~\cite{sartor:1980,grange:1987,jaminon,baldo:1990,baldo:1991,mahaux:1993,hassaneen:2004},
the in-medium T-matrix approach~\cite{bozek:2002,soma:2008}, the variational Monte Carlo method~\cite{schiavilla:2007},
and the correlated basis function approach~\cite{fantoni:1984,benhar:1990}.
 The predicted depletion of the Fermi sea
by different theoretical approaches have been shown to a slightly larger than
15\%~\cite{grange:1987,baldo:1991,fantoni:1984,benhar:1990}, in agreement with the experimental observations.
In Ref.\cite{baldo:1991}, the nucleon momentum distribution has been studied within the Brueckner-Bethe-Goldstone theory
by including high-order contributions in the hole-line expansion of the mass operator. In our previous paper\cite{yin:2013},
the effect of a microscopic three-body force (TBF) on the nucleon momentum distribution has been discussed and it is shown that
inclusion of the TBF may lead to a enhancement of the depletion of nuclear Fermi sea in nuclear matter at high densities
well above the saturation density.
 In the present paper, we shall
investigate the effect of tensor correlations on the depletion of nuclear Fermi sea within the
framework of the extended BHF approach.


\section{Theoretical Approaches}
The present calculations are based on the extended BHF approach~\cite{jeukenne:1976}.
The extension of the Brueckner approach to include microscopic TBFs
can be found in Ref.\cite{grange:1989,zuo:2002a}.
The starting point of the BHF approach is the reaction $G$-matrix, which satisfies the
Bethe-Goldstone (BG) equation~\cite{day},
\begin{eqnarray}\label{eq:BG}
G(\rho;\omega)= V+
 V\sum\limits_{k_{1}k_{2}}\frac{|k_{1}k_{2}\rangle
Q(k_{1},k_{2})\langle
k_{1}k_{2}|}{\omega-\epsilon(k_{1})-\epsilon(k_{2})}G(\rho;\omega)
\end{eqnarray}
where $k_i\equiv(\vec k_i,\sigma_1,\tau_i)$ denotes the momentum,
the $z$-component of spin and isospin of a nucleon, respectively;
$V$ is the realistic $NN$ interaction; $\omega$ is starting energy; $Q(k_{1},k_{2})$ is the
the Pauli operator. The single-particle (s.p.) energy $\epsilon(k)$ is given by:
$\epsilon(k)=\hbar^{2}k^{2}/(2m)+U_{\rm BHF}(k)$.
We adopt the continuous choice for
the auxiliary potential $U_{\rm BHF} (k)$ in the present calculation.
Under the continuous choice, the s.p. potential describes physically at the lowest BHF level
the nuclear mean field felt by a nucleon in nuclear medium and is
calculated as follows:
\begin{eqnarray}\label{eq:UBHF}
U_{\rm BHF}(k)=Re\sum\limits_{k'\leq k_{F}}\langle
kk'|G[\rho,\epsilon(k)+\epsilon(k')]|kk'\rangle_{A}.
\end{eqnarray}

For the realistic $NN$ interaction, we adopt the
$AV18$ two-body interaction~\cite{wiringa:1995} plus a
microscopic TBF~\cite{zuo:2002a} constructed by using the meson-exchange current
approach~\cite{grange:1989}.
The parameters of the TBF model, i.e., the coupling constants and the form factors,
have been self-consistently determined to reproduce the $AV18$
two-body force using the one-boson-exchange potential
model and their values can be found in Ref.\cite{zuo:2002a}.
 In our calculation, the TBF
contribution has been included by reducing the TBF to an
equivalently effective two-body interaction according to the
standard scheme as described in Ref.\cite{grange:1989}.

In order to calculate the nucleon momentum distribution in nuclear matter with the EBHF approach,
we follow the scheme given in Refs.~\cite{jeukenne:1976,baldo:1991}.
 Within the framework of the Brueckner-Bethe-Goldstone theory, the mass operator can be expanded in a perturbation series
according to the number of hole lines, i.e.,
\begin{equation}
 M (k,\omega) = M_1(k,\omega)+M_2(k,\omega)+M_3(k,\omega)+\cdots
 \end{equation}
The mass operator is complex quantity and the real part of its on-shell value
can be identified with the potential energy felt by a neutron or a proton in nuclear matter.
In the expansion of the mass operator, the first-order contribution
$M_1(k,\omega)$ corresponds to the standard BHF s.p. potential and the real part of its on-shell value coincides
with the auxiliary potential under the continuous choice given by Eq.~(\ref{eq:UBHF}).
The higher-order terms stems from
the density dependence of the effective $G$-matrix, which are important for describing realistically and reliably
the s.p. properties within the Brueckner theory~\cite{jeukenne:1976}.
The second-order term $M_2$ is called Pauli rearrangement term and it describes the effect of ground
state correlations \cite{baldo:1988,baldo:1990}.
The effect of ground
state correlations is not only essential for getting a satisfactorily agreement between
the predicted depth of the microscopic BHF s.p. potential and the empirical
value~\cite{jeukenne:1976} and for restoring the restoring the Hugenholtz-Van Hove
theorem \cite{zuo:1999},
but also plays a crucial role in generating a nucleon self-energy to describes
realistically the s.p. strength distribution in nuclear matter and finite nuclei below the
Fermi energy~\cite{dickhoff:1992}. The Pauli rearrangement contribution can be
calculated according to Ref.\cite{jeukenne:1976}.
The third-order term $M_3$ is called renormalization contribution and it takes
into account the above effect of the depletion of the Fermi seas.
The renormalization term $M_3$ is given by~\cite{jeukenne:1976,baldo:1990,zuo:1999}:
\begin{equation}
M_3(k,\omega)= - \sum_{h}\kappa_2(h)
\langle kh|G(\omega + \epsilon(h))|kh \rangle_A \ ,
\label{eq:M3}
\end{equation}
where $h$ refers to the hole state below the Fermi momentum, and
$\kappa_2(h)=-\left[ \partial M_1(h,\omega) / {\partial\omega} \right]_{\omega=\epsilon(h)}$
is the depletion of the Fermi sea at the lowest-order approximation \cite{jeukenne:1976,zuo:1999}.
As shown in Ref.~\cite{baldo:1991}, it is a satisfactory
approximation to replace the depletion coefficient $\kappa_2(h)$ in Eq.~(\ref{eq:M3}) by its value
at the averaged momentum inside the Fermi sea, i.e., $\kappa =\kappa_2(h=0.75k_F)$.
By taking into account $M_3(k,\omega)$, one gets the {\it renormalized} BHF
approximation for the mass operator~\cite{jeukenne:1976}, i.e.,
\begin{equation}
 \widetilde{M}_1(k,\omega) \equiv M_1(k,\omega)+M_3(k,\omega) \approx
( 1 - \kappa ) M_1(k,\omega) \ .
\end{equation}
Similarly one may obtain the {\it renormalized} $\widetilde{M}_2$ \cite{jeukenne:1976}:
$\widetilde{M}_2(k,\omega)= (1 - \kappa) M_2(k,\omega)$.
In terms of the off-shell mass operator, it is readily to calculate the momentum distribtution below and above
the Fermi momentum \cite{jeukenne:1976,baldo:1991}:
\begin{eqnarray}
n(k) = 1+ \left[\partial {\widetilde{U}_1(k,\omega)} / \partial \omega
\right]_{\omega=\epsilon(k)}, \ \ \ \ \ \  {\rm for} \ k < k_F
\\
n(k) = - \left[ \partial {\widetilde{U}_2(k,\omega)} / \partial \omega
\right]_{\omega=\epsilon (k)}, \ \ \ \ \ \  \ \ \ \ \ {\rm for} \ k> k_F
\end{eqnarray}
where $\widetilde{U}_1$ and $\widetilde{U}_2$ denote the real parts of
$\widetilde{M}_1$ and $\widetilde{M}_2$, respectively.

\section{Results and Discussions}

\begin{figure}[tbh]
\centerline{\includegraphics[width=0.5\textwidth]{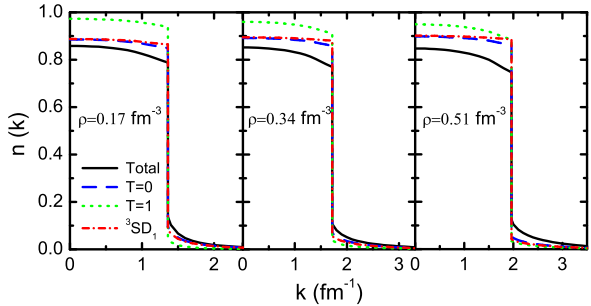}}
\caption{(Color online) Contributions from different partial wave channels to nucleon momentum distribution
in symmetric nuclear matter at three typical
densities.
}\label{fig1}
\end{figure}
\begin{table}[ht]
\caption{The calculated contributions from
different channels to the depletion [i.e., $1-n(k=0)$] of the lowest hole state in nuclear matter.
The results in this table have been obtained without including the TBF. }\label{tab1}
\begin{ruledtabular}
\begin{tabular}{lllll}
 $\rho$ (fm$^{-3}$)  & $T=1$  & $T=0$ & $^3SD_1$ & total  \\
\hline
&&&& \\
0.17 & 0.027 & 0.115 & 0.113 & 0.142  \\
0.34 & 0.040 & 0.108 & 0.105 & 0.148 \\
0.51 & 0.051 & 0.102 & 0.099 & 0.153 \\
\end{tabular}
\end{ruledtabular}
\end{table}
In Fig.\ref{fig1} we show the contributions from different partial wave channels to the nucleon momentum distribution
in symmetric nuclear matter at three typical
densities $\rho=0.17, 0.34$ and 0.51fm$^{-3}$, respectively.
In the figure, the solid lines
correspond to the total distributions; the dashed ones denotes the contribution from the isospin $T=0$ $np$ channels; the
 dotted curves are the contribution from the isospin-triplet $T=1$ channels; the dot-dashed curves indicate the contribution
 from the $T=0$ $^3SD_1$ tensor channel. The results of Fig.\ref{fig1} are obtained by adopting purely the $AV18$ two-body interaction
 and the TBF is not included.
 It is seen that the $NN$ correlations may lead to
 a depletion of the nucleon hole states below the Fermi surface and
 a partly occupation of the particle states above the Fermi surface in the correlated ground state of nuclear matter.
We notice that the depletion of the Fermi sea is mostly induced by the isospin $T=0$ $np$ correlations.
In Tab.\ref{tab1}, we report the predicted contributions from
different channels to the depletion [i.e., $1-n(k=0)$] of the lowest momentum state in nuclear matter.
It is seen from the Tab.\ref{tab1} that at the lowest momentum ($k=0$), the $T=0$ channels contribute a $11.5\%$ ($10.8\%$, $10.2\%$) depletion
which amounts to almost 81, 73, 67 percent to the total $14.2\%$, $14.8\%$, $15.3\%$ depletion at $\rho=0.17$fm$^{-3}$, $\rho=0.34$fm$^{-3}$, and $\rho=0.51$fm$^{-3}$), respectively.
Whereas, the contribution from the $T=1$ channels amounts
only about 19, 27 and 33 percent of the total depletion for
densities $\rho=0.17$, 0.34 and 0.51fm$^{-3}$, respectively.
It is worth noticing from Fig.~\ref{fig1} that the dashed (red) curves are almost overlap with the corresponding dot-dashed (blue) curves except for momenta around the Fermi momentum, indicating that the contribution of the $T=0$ $np$ channel almost completely stems from the $SD$ tensor correlations.  Therefore, we may conclude that the depletion of the Fermi sea in symmetric nuclear matter is essentially dominated by the $SD$ $np$ tensor correlations.
\begin{figure}[tbh]
\centerline{\includegraphics[width=0.5\textwidth]{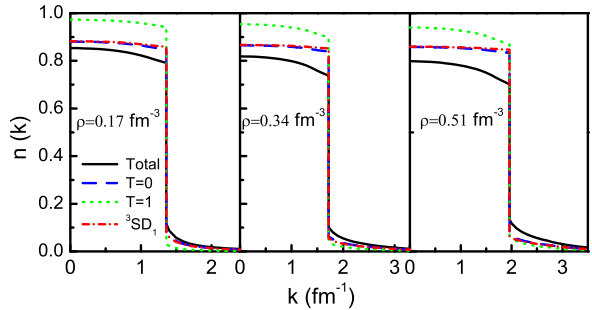}}
\caption{(Color online) The same as Fig.~\ref{fig1}, but obtained by including the TBF.}
\label{fig2}
\end{figure}
\begin{table}[ht]
\caption{The same as TABLE \ref{tab1}, but the results are obtained by including the TBF.}\label{tab2}
\begin{ruledtabular}
\begin{tabular}{lllll}
 $\rho$ (fm$^{-3}$)  & $T=1$  & $T=0$ & $^3SD_1$ & total  \\
\hline
&&&& \\
0.17 & 0.028 & 0.119 & 0.118 & 0.147  \\
0.34 & 0.046 & 0.135 & 0.133 & 0.181 \\
0.51 & 0.060 & 0.142 & 0.140 & 0.202 \\

\end{tabular}
\end{ruledtabular}
\end{table}

To see the TBF effect, we show in Fig.~\ref{fig2} the same results as in Fig.~\ref{fig1} but obtained by adopting the $AV18$
two-body interaction plus the microscopic TBF. As already shown in Ref.~\cite{yin:2013} that the TBF effect on the nucleon
momentum distribution at relative low densities around the empirical saturation density is negligibly small. Whereas, the TBF effect
becomes noticeable at high densities well above the saturation density and it has been shown to enhance the depletion of the nuclear
Fermi sea~\cite{yin:2013}.
The TBF-induced enhancement of the depletion turns out to be more pronounced at higher densities since
the TBF is expected to induce stronger short-range correlations at higher densities. Tab.\ref{tab2} gives the contributions from
different channels to the depletion of the lowest momentum state predicted by including the TBF. By comparing Tab.\ref{tab1} and Tab.\ref{tab2},
one may notice that inclusion of the TBF increases the depletion of the
lowest hole state from $14.2\%$, $14.8\%$ and $15.3\%$ to $14.7\%$, $18.1\%$ and $20.2\%$ for $\rho=0.17, 0.34$ and 0.51fm$^{-3}$, respectively.
One may notice that inclusion of the TBF does not alter the main conclusions obtained in the case of not including the TBF, i.e., the
depletion of nuclear Fermi sea is dominated by the $T=0$ $SD$ tensor correlations between neutron and proton.
For the lowest momentum state, the $T=0$ channels contribute a $11.9\%$ ($13.5\%$, $14.2\%$) depletion
which amounts to almost 81 (75, 70) percent to the total $14.7\%$ ($18.1\%$, $20.2\%$) depletion at $\rho=0.17$fm$^{-3}$ ($\rho=0.34$fm$^{-3}$, $\rho=0.51$fm$^{-3}$).
The isospin-singlet $T=0$ channels contain purely the $np$ correlations since
the $nn$ and $pp$ interactions vanish in the $T=0$ channels. Whereas, the isospin-triplet $T=1$ channels consist three $T_z$ components, i.e., $T_z=-1,0$ and 1,
and consequently contain both $np$ and $nn$ ($pp$) correlations. If neglecting the small charge-breaking effect in the $AV18$ interaction, we may expect the
$nn$, $np$, $pn$ and $pp$ correlations\footnote{We distinguish explicitly the notation $np$ and $pn$ only in this sentence and the following sentence simply for convenience of our discussion.} in the $T=1$ channels are approximately of the same strength. Accordingly, the $nn$, $np$, $pn$ and $pp$ correlations contribute equally to the depletion in the $T=0$ channels and
each accounts for $25\%$. Therefore, the depletion due to the $nn$ or $pp$ correlations
can be estimated to amount to about $4.8\%$, $6.3\%$ and $7.5\%$ out of the total depletion of $14.7\%$, $18.1\%$ and $20.2\%$
at $\rho=0.17$, 0.34, 0.51fm$^{-3}$, respectively.
Finally, we arrive the conclusion that about $90.6\%$ ($87.6\%$, $85.0\%$) percent of the total depletion of the lowest hole state
stems from the $np$ correlations,
whereas only about $4.8\%$ ($6.3\%$, $7.5\%$) is induced by the $nn$ or $pp$ correlations for $\rho=0.17$fm$^{-3}$ (0.34, 0.51fm$^{-3}$).
In Ref.~\cite{subedi:2008}, the definite experimental evidence for
the strong enhancement of the $np$ short-range correlations
over the $pp$ and $nn$ correlations observed at JLab has been reported. It is shown in Ref.~\cite{subedi:2008} that
$80\%$ of the nucleons in the $^{12}C$ nucleus acted independently or as described
within the shell model, whereas among the
$20\%$ of correlated pairs, $90\pm 10\%$ are in the
form of $pn$ short-range correlated ($SRC$) pairs; only $5\pm 1.5\%$ are in the form
of $pp$ $SRC$ pairs and $5\pm 1.5\%$ in the $nn$ $SRC$ pairs.
Our above results provide a microscopic support
for the experimental observation at Jlab.

\begin{figure}[tbh]
\centerline{\includegraphics[width=0.5\textwidth]{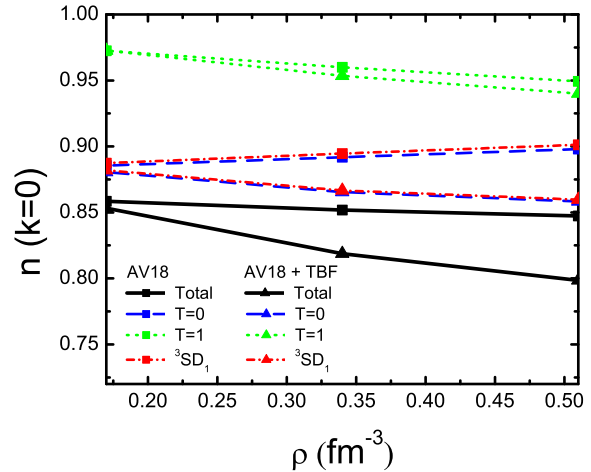}}
\caption{ (Color online) The density dependence of the occupation probabilities of the lowest hole state
predicted by considering the $T=0$, $T=1$, $^3SD_1$ channel correlations and the total correlations, respectively.
The square symbols are obtained without the TBF; the triangles are calculated by including the TBF.} \label{fig3}
\end{figure}

In Fig.~\ref{fig3}, we display the predicted density dependence of the occupation probabilities of the lowest hole state in nuclear matter.
It can be seen that without the TBF, the total depletion is almost independent of density
in the region from $0.17$fm$^{-3}$ to $0.51$fm$^{-3}$, in good agrement with the previous BHF result \cite{baldo:1991}
and the prediction of Ref.~\cite{rios:2009} by using the Green Function theory.
In the case of excluding the TBF, the depletion due to the $T=1$ correlations increases and that caused by the $T=0$ correlations increases
slowly as the density increases. The TBF effect on the $T=0$ channel depletion is seen to be quite weak and it leads to a slight
enhancement at high densities. Whereas, inclusion of the TBF enhances considerably the depletion induced by the $T=0$
correlations at high densities well above the saturation density. One may see that the TBF-induced enhancement of the total depletion
and the $T=0$ channel depletion increases monotonically as a function of density. Again it is worth noticing that the $T=0$ $np$ correlations
play a predominated role in generating the depletion over the $T=1$ correlations in the whole density range considered here.
The $T=0$ channel depletion turns out to determined almost completely by the $^3SD_1$ tensor correlations.

\section{summary}
In summary, we have investigated the effect of tensor correlations on the depletion of nuclear Fermi sea in nuclear matter
within the framework of the extended BHF approach by adopting the $AV18$ two-body interaction supplemented with a microscopic TBF.
The contributions from various partial wave channels including
the isospin-singlet $T=0$ channels, the isospin-triplet $T=1$ channels and the $T=0$ tensor $^3 SD _1$ channel have been
calculated and discussed.
The main conclusions can be summarized as follows:
(1) The $T=0$ neutron-proton correlations play a predominated role in generating the depletion
of nuclear Fermi sea in symmetric nuclear matter over the $T=1$ correlations,
which provide a robust and microscopic support for the recent experimental
observation at Jlab \cite{subedi:2008}. (2) The $T=0$ correlation-induced depletion stems almost completely from the $^3SD_1$ tensor channel.
(3) The TBF effect on the depletion is shown to increases monotonically as a
 function of density and it enhances sizably the total depletion and the depletion caused by the $T=0$ $np$ correlations at high densities well
 above the saturation density.

\section*{Acknowledgments}

{The work is supported by the National Natural Science
Foundation of China (11175219), the 973 Program of China (No.2013CB834405), the
Knowledge Innovation Project (KJCX2-EW-N01) of Chinese Academy of
Sciences, and the Project of Knowledge Innovation Program
(PKIP) of Chinese Academy of Sciences, Grant No. KJCX2.YW.W10.}


\begin{thebibliography}{00}

\bibitem{otsuka:2005} T. Otsuka, T. Suzuki, R. Fujimoto1, H. Grawe, and Y. Akaishi,
Phys. Rev. Lett. {\bf 95}, 232502 (2005).

\bibitem{otsuka:2010} T. Otsuka, T. Suzuki, M. Honma, Y. Utsuno, N. Tsunoda, K. Tsukiyama, and M. Hjorth-Jensen,
  Phys. Rev. Lett. {\bf 104}, 012501 (2010).

\bibitem{vidana:2011} I. Vidana, A. Polls and C. Providencia,
Phys. Rev. {\bf C 84}, 062801 (2011).

\bibitem{frankfurt:2008} L. Frankfurt, M. Sargsian, and M. Strikman, Int. J. Mod. Phys.
{\bf A 23}, 2991 (2008).

\bibitem{jeukenne:1976} J. P. Jeukenne, A. Lejeune and C. Mahaux, Phys. Rep. {\bf 25}, 83 (1976).

\bibitem{vonderfecht:1991} B. Vonderfecht, W. Dickhoff, A. Polls and A. Ramos, Phys. Rev. {\bf C44}, R1265 (1991).

\bibitem{cavedon:1982} J. M. Cavedon, B. Frois, D. Goutte, {\it et al.}, Phys. Rev. Lett. {\bf 49}, 978 (1982).

\bibitem{pandharipande:1997} V. R. Pandharipande, I. Sick and P. K. A. deWitt Huberts, Rev. Mod. Phys. {\bf 69}, 981 (1997).

\bibitem{ramos:1989} A. Ramos, A. Polls, and W. H. Dickhoff, Nucl. Phys. {\bf A503}, 1 (1989).

\bibitem{dickhoff:1992}W. H. Dickhoff and M. M\"uther, Rep. Prog. Phys. {\bf 55}, 1947 (1992).

\bibitem{dickhoff:2004} W. Dickhoff and C. Barbieri,  Prog. Part. Nucl. Phys. {\bf 52}, 377 (2004).


\bibitem{mitt:1990} P. K. A. deWitt Huberts, J. Phys. {\bf G16}, 507 (1990).

\bibitem{lapikas:1999} L. Lapik\'{a}s, J. Wesseling, and R. B. Wiringa, Phys. Rev. Lett. {\bf 82}, 4404 (1999).

\bibitem{starink:2000}  R. Starink, M. F. van Batenburg, E. Cisbani {\it et al.}, Phys. Lett. {\bf B474}, 33 (2000).

\bibitem{batenburg:2001} M. F. van Batenburg, Ph. D. thesis, University of Utrecht, 2001 (unpublished).

\bibitem{rohe:2004} D. Rohe {\it et al.} (E97-006 Collaboration), Phys. Rev. Lett. {\bf 93}, 182501 (2004).


\bibitem{niyazov:2004} R. A. Niyazov {\it et al.} (CLAS Collaboration), Phys. Rev. Lett. {\bf 92}, 052303 (2004);
K. S. Egiyan {\it et al.} (CLAS Collaboration), Phys. Rev. Lett. {\bf 96}, 082501 (2006).

\bibitem{benmokhtar:2005} F. Benmokhtar {\it et al.} (Jefferson Lab Hall A Collaboration), Phys. Rev. Lett. {\bf 94}, 082305 (2005);
R. Shneor {\it et al.} (Jefferson Lab Hall A Collaboration), Phys. Rev. Lett. {\bf 99}, 072501 (2007).


\bibitem{aclander:1999} J. L. S. Aclander, J. Alster, D. Barton {\it et al.}, Phys. Lett. {\bf B 453}, 211 (1999);
A. Tang, J. W. Watson, J. Aclander {\it et al.}, Phys. Rev. Lett. {\bf 90}, 042301 (2003);
E. Piasetzky, M. Sargsian, L. Frankfurt, M. Strikman, and J. W. Watson, Phys. Rev. Lett. {\bf 97}, 162504 (2006).

\bibitem{onderwater:1998} C. J. G. Onderwater, K. Allaart, E. C. Aschenauer {\it et al.}, Phys. Rev. Lett. {\bf 81}, 2213
(1998).

\bibitem{reley:2008} L. A. Riley, P. Adrich, T. R. Baugher, {\it et al.} Phys. Rev. {\bf C 78}, 011303 (2008).

 \bibitem{subedi:2008} R. Subedi, R. Shneor, P. Monaghan, {\it et al.}, Science {\bf 320}, 1476 (2008) and reference therein.


\bibitem{schiavilla:2007} R. Schiavilla, R. B. Wiringa, S. C. Pieper, and J. Carlson, Phys.
Rev. Lett. {\bf 98}, 132501 (2007).

\bibitem{muther:1995} H. M\"{u}ther, G. Knehr and A. Polls, Phys. Rev. {\bf C 52}, 2955 (1995).


\bibitem{alm:1996} T. Alm, G. R\"opke, A. Schnell, N. H. Kwong, and H. S. K\"ohler,
Phys. Rev. {\bf C53}, 2181 (1996).

\bibitem{dewulf:2003}
 Y. Dewulf, D. Van Neck, and M. Waroquier,
 Phys. Rev. {\bf C65}, 054316 (2002);
Y. Dewulf, W. H. Dickhoff, D. Van Neck, E. E. Stoddard, and M.
Waroquier, Phys. Rev. Lett. {\bf 90}, 152501 (2003).

 \bibitem{frick:2005} T. Frick, H. M\"uther, A. Rios, A. Polls, and A. Ramos,
 Phys. Rev. {\bf C 71}, 014313 (2005).

\bibitem{rios:2009} A. Rios, A. Polls, I. Vidana, Phys. Rev. C 79, 025802 (2009);
A. Rios, A. Polls and W. Dickhoff, Phys. Rev. {\bf C79}, 064308 (2009).


 \bibitem{sartor:1980} R. Sartor and C. Mahaux, Phys. Rev. {\bf C 21}, 1546 (1980).

\bibitem{grange:1987} P. Grang\'{e}, J. Cugnon, and A. Lejeune, Nucl. Phys. {\bf A473}, 365 (1987).

 \bibitem{jaminon} M. Jaminon and C. Mahaux, Phys. Rev. {\bf C41}, 697 (1990).

\bibitem{baldo:1990} M. Baldo, I. Bombaci, G. Giansiracusa, U. Lombardo, C. Mahaux and R. Sartor,
  Phys. Rev. {\bf C41}, 1748 (1990).

\bibitem{baldo:1991} M. Baldo, I. Bombaci, G. Giansiracusa and U. Lombardo, Nucl. Phys. {\bf A530}, 135 (1991).

\bibitem{mahaux:1993} C. Mahaux and R. Sartor, Nucl. Phys. {\bf A553}, 515 (1993).

\bibitem{hassaneen:2004} Kh. S. A. Hassaneen and H. M\"{u}ther, Phys. Rev. {\bf C 70}, 054308 (2004)




\bibitem{bozek:2002} P. Bozek, Phys. Rev. {\bf C 59}, 2619 (1999);
P. Bozek, Phys. Rev. {\bf C65}, 054306 (2002).
\bibitem{soma:2008} V. Soma and P. Bozek, Phys. Rev. {\bf C 78}, 054003 (2008)

\bibitem{fantoni:1984} S. Fantoni and V. R. Pandharipande, Nucl. Phys. {\bf A427}, 473 (1984).

\bibitem{benhar:1990} O. Benhar, A. Fabrocini, and S. Fantoni, Nucl. Phys. {\bf A505}, 267 (1989);
O. Benhar, A. Fabrocini, and S. Fantoni, Phys. Rev. {\bf C 41}, R24
(1990).


\bibitem{yin:2013} Yin Peng, Jian-Yang Li, Pei Wang and Wei Zuo, Phys. Rev. {\bf C }, (2013).

\bibitem{grange:1989} P.~Grang\'e, A.~Lejeune, M.~Martzolff, and J.-F.Mathiot,
 Phys. Rev. {\bf C 40}, 1040(1989).

\bibitem{zuo:2002a} W. Zuo, A. Lejeune, U. Lombardo {\it et al.},
Nucl. Phys. {\bf A 706}, 418 (2002); Eur. Phys. J. {\bf A 14}, 469 (2002).

\bibitem{day} B. D. Day, Rev. Mod. Phys. {\bf 50}, 495 (1978).



\bibitem{wiringa:1995} R. B. Wiringa, V. G. J. Stoks and R. Schiavilla,
{\it Phys. Rev.} {\bf C 51}, 38 (1995).

\bibitem{baldo:1988} M. Baldo, I. Bombaci, L. S. Ferreira, G.
Giansiracusa, U. Lombardo, Phys. Lett. {\bf B 209}, 135 (1988);
Phys. Lett. {\bf B 215}, 19 (1988).

\bibitem{zuo:1999} W. Zuo, I. Bombaci and U. Lombardo, Phys. Rev. {\bf C60}, 024605 (1999).












\end{thebibliography}
\end{document}